\documentclass[a4paper,twocolumn,10pt]{article}
\usepackage{eurosis}
\usepackage{url}
\usepackage{natbib}
\usepackage{amsmath}
\usepackage{mathtools}
\usepackage{mathptmx}
\usepackage{caption}
\usepackage{calrsfs}
\usepackage{multirow}
\usepackage{orcidlink}
\hypersetup{hidelinks}
\usepackage{bbding}
\DeclareMathAlphabet{\pazocal}{OMS}{zplm}{m}{n}

\bibpunct[; ]{(}{)}{,}{a}{}{;}

\begin{document}

%
%
\title{\MakeUppercase{Capacitive Touch Sensor Modeling with a Physics-informed Neural Network and Maxwell's Equations}}


\author{Ganyong Mo$^{1,2} \orcidlink{0009-0002-0653-2203}$, Krishna Kumar Narayanan$^{1} \orcidlink{0009-0008-6437-7145}$, David Castells-Rufas$^{2} \orcidlink{0000-0002-7181-9705}$, Jordi Carrabina$^{2} \orcidlink{0000-0002-9540-8759}$\\
$^1$ KOSTAL Electrica, S.A., R\&D Department, Polinyà, Spain, \texttt{FirstLetterFirstName.LastName@kostal.com}\\
$^2$ Univ. Autònoma de Barcelona, Microelectronic\&Electronic System, Bellaterra, Spain, \texttt{FirstName.LastName@uab.cat}\\
}


\date{}

\maketitle

\thispagestyle{empty}


\begin{abstract}
Maxwell's equations are the fundamental equations for understanding electric and magnetic field interactions and play a crucial role in designing and optimizing sensor systems like capacitive touch sensors, which are widely prevalent in automotive switches and smartphones. Ensuring robust functionality and stability of the sensors in dynamic environments necessitates profound domain expertise and computationally intensive multi-physics simulations. This paper introduces a novel approach using a Physics-Informed Neural Network (PINN) based surrogate model to accelerate the design process. The PINN model solves the governing electrostatic equations describing the interaction between a finger and a capacitive sensor. Inputs include spatial coordinates from a 3D domain encompassing the finger, sensor, and PCB, along with finger distances. By incorporating the electrostatic equations directly into the neural network's loss function, the model captures the underlying physics. The learned model thus serves as a surrogate sensor model on which inference can be carried out in seconds for different experimental setups without the need to run simulations. Efficacy results evaluated on unseen test cases demonstrate the significant potential of PINNs in accelerating the development and design optimization of capacitive touch sensors. 

\end{abstract}

\section{\MakeUppercase{Introduction}}

\begin{figure*}[ht]
    \centering
    \includegraphics[width=0.95\textwidth]{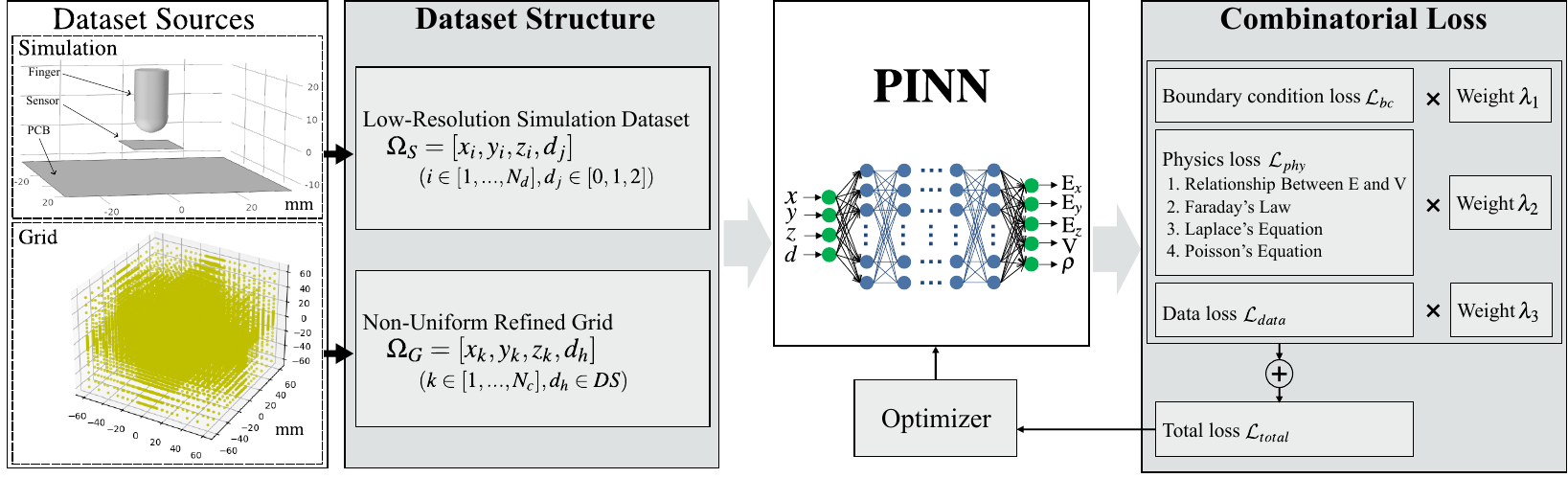}
    \caption{System Architecture of the PINN Process}
    \label{fig: systematic architecture}
\end{figure*}

Capacitive sensors with no mechanical parts are fast becoming the preferred human-machine interface (HMI) within the automotive sector. 
The versatility and flexibility of utilizing the sensor foil across curved surfaces have enabled the development of a wide range of creative designs and innovative user interfaces. They can be found on a variety of devices in an automobile, ranging from switches, gauges, buttons, seats, steering wheels, etc., with touch, gesture, and hover being typical user interactions. For the reliable functioning of such a variety of applications, one must ensure that the sensors satisfy robust design constraints and environmental factors. Temperature, humidity, and electromagnetic interference (EMI) significantly influence the dielectric properties of the sensor and introduce noise to the signals, leading to performance deterioration, thereby affecting the long-term stability and accelerating sensor degradation.

Current industry solutions are addressed through a combination of compensation models \citep{gebhardt2006temperature} and denoising filter mechanics. Sensor models are simulated using different multi-physics simulation tools like COMSOL and Ansys, which use finite element method (FEM) to model the electrostatic interaction between the sensor, environment, and finger. But FEM simulations work by discretizing the domain into finite elements and solving equations at every single node. This requires the generation of a fixed-resolution mesh, and it is challenging when the experimental setup changes, such as the geometry of the sensor and finger. 

Deep learning (DL) methods provide an alternative approach where a surrogate sensor model is trained from a large set of input-output data. However, such networks lack physical constraints and fail to accurately capture behavior when presented with novel data. PINNs \citep{raissi2019physics} is a rapidly growing sub-field of DL that addresses this problem by integrating the governing equations and the system constraints directly into the neural network (NN) architecture. In the case of sensor models, compared with the FEM simulation, PINNs can offer a continuous approximation of the solution over the entire domain using NNs. This allows for a multi-resolution output without interpolation. Furthermore, PINNs are mesh-free and adapt easily to changing conditions, thus providing an ideal base to perform fast inference on a variety of setups for establishing the functional performance of a specific setup. This makes PINNs an ideal tool for performing generative design and optimization. We take motivation from this, where we use PINNs to obtain fast and accurate estimations of the electrostatic characteristics for different finger distances, thereby providing the bedrock for future studies regarding the long-term performance of a capacitive sensor.

In this paper, we intend to make the initial stride towards this goal by training NN to capture the electrostatic interaction between a capacitive sensor and a finger. The outputs of the network are the electric field distribution ($\mathrm{E}$), electric potential ($\mathrm{V}$), and charge density ($\rho$). We utilize an electroquasistatic approximation of Maxwell's equations, which ignores the magnetic fields with an insignificant loss of accuracy \citep{1996CapacitiveSD}. Given a single array capacitive sensor with fixed dimensions and a finger modeled as a cylinder with a rounded end, the variables $\mathrm{E}$, $\mathrm{V}$, and $\mathrm{\rho}$ are predicted for different distances ($d$) between the sensor and the finger. A coarse-resolution dataset at three specific distances is generated from a multi-physics simulation. Loss computed from this sparse data provides the initial estimate for the PINN which is then complemented with the physics residual losses arising from the governing partial differential equations (PDEs). This loss acts as a regularization agent to limit the admissible solution space and captures the right solution even at distances and locations not presented in the training data. 
The authors of \citep{lim2022maxwellnet} train a NN that solves Maxwell's equations to predict the electric field given a 2D sample shape of various aspheric optical lenses. The authors of \cite{Pan_2020} solve the time domain electromagnetic simulation based on Maxwell's equations on a 1D cavity model filled with homogeneous media using an unsupervised PINN. Our setup differs in that we solve Maxwell's equations in the 3D domain, which increases the complexity of the problem space and computational requirements. Hence, in order to speed up the convergence and guarantee the quality of model training, we employ the combinatorial loss structure \citep{ma2020combined} involving both data-driven and physics-driven loss. While it is theoretically possible to train PINNs without any data, incorporating data can enhance convergence and guide the learning process more effectively. Additionally, by using only a sparse dataset acquired from the commercial computational solver, we reduce the computational load and make the training process faster without any loss of accuracy. Figure \ref{fig: systematic architecture} shows the system architecture of the overall PINN process used. The inputs to the NNs are the spatial coordinates and the finger distances. The coordinates are aggregated from two 3D sources: 1. a low-resolution electrostatic simulation dataset $\Omega_S$ at three fixed finger distances $d_{j}=[d_0,d_1,d_2]$ from the sensor; and 2. collocation points sampled from a refined non-uniform 3D grid $\Omega_G$. The outputs of the NN are the electric field distributions ($\mathrm{E}_x,\mathrm{E}_y,\mathrm{E}_z$), the electric potential ($\mathrm{V}$), and the charge density ($\rho$). The total loss is composed of a data-driven loss computed only on $\Omega_S$ and physics-driven losses computed over the collocation points sampled from $(\Omega_{S} + \Omega_G)$. 







The paper is organized as follows: In the section \textit{methodology}, the governing Maxwell's equations capturing the electrostatic characteristics are presented, followed by the NN setup, the loss function, and performance metrics. The NN results on unseen datasets and their generalization ability are shown in the section \textit{experiments and results}. Finally, the concluding remarks and the outlook are provided in the section \textit{conclusion}.

\section{\MakeUppercase{Methodology}}\label{sec:methodology}
In this section, the proposed methodology for setting up the PINN framework to learn 3D electrostatic Maxwell's equations is presented. 
\paragraph{Governing Equations}
Maxwell's equations relating the electric and magnetic fields, charge density, and current density form the governing physics to describe the fields and currents that make up the capacitive sensor. In our case, the magnetic fields are ignored due to the relatively low-frequency regime \citep{1996CapacitiveSD}. 
\begin{align}
    \nabla \times \text{E} &= \frac{\partial}{\partial t} \mu_0 \mathrm{H} = 0 \label{eq:faraday} \\
    \nabla \cdot \text{D} &= \rho  \label{eq:gauss}
\end{align}
Eq. \ref{eq:faraday} is the Faraday's law from Maxwell's equations, where $\mu_0$ refers to the magnetic permittivity of vacuum and $\mathrm{E}$ is a continuous distribution vector field that is defined for any given points in the interelectrode space between the conductors and is composed of three components ($\mathrm{E}_x, \mathrm{E}_y, \mathrm{E}_z$). The curl ($\nabla \times$) of $\mathrm{E}$ is set to zero as the magnetic field intensity ($\mathrm{H}$) is ignored in our case and approximated by zero. On the other hand, Eq. \ref{eq:gauss} shows the Gauss's law from Maxwell's equations, which defines the electric charge density $\rho$ as the divergence ($\nabla \cdot)$ of the electric displacement field $\mathrm{D}$ on the conductor. Thus, Maxwell's equations reduce to their electroquasistatic form and are solved for every input spatial coordinate $\mathrm{X}=(x,y,z)$ presented to the neural network. In a homogenous isotropic medium with no free charge, $\rho$ is equal to zero, thus giving us the Laplace equation. The charges on the equipotent surface conductors (sensor and PCB) are distributed among themselves, and thus the Laplace equation is satisfied in the interelectrode space between the conductors, just like Eq. \ref{eq:faraday}. In areas outside of the interelectrode space, if we approximate the shape of the sensor and PCB as 2D surfaces with negligible thickness, one can simplify the differential form of volume charge density Eq. \ref{eq:gauss} to a surface charge density ($\rho_s$) given by: 
\begin{equation}\label{eq:Gauss_nd}
\mathrm{n} \cdot \mathrm{D} = \rho_s 
\end{equation}
where $\mathrm{n}$ is the surface normal of the PCB and sensor. The shape simplification thus presents us with an essential boundary condition (BC) where the surface charge at the material interface is equated to the jump in the displacement field along its surface normal component. 
To obtain a closed form, the constitutive relation between $\mathrm{D}$ and $\mathrm{E}$ is utilized (Eq.\ref{eq:const_rel}). 
\begin{equation}
    \label{eq:const_rel}
    \mathrm{D} = \varepsilon_0 \mathrm{E} + \mathrm{P}
\end{equation}
$\mathrm{P}$ is the polarization density, and $\varepsilon_0$ is the vacuum permittivity. Substituting the polarization for linear materials $\mathrm{P}=\varepsilon_0 \chi_{e} \mathrm{E}$ with $\chi_{e}$ representing the electric susceptibility in Eq. \ref{eq:const_rel}, we obtain the simplified formulation of the constitutive relation:
\begin{equation}
	\label{eq:simp_const_rel}
    \mathrm{D} = \varepsilon_0 (1+\chi_e)\mathrm{E} = \varepsilon_0 \varepsilon_r \mathrm{E}
\end{equation}
$\varepsilon_r$ refers to the relative permittivity of the material. Thus, from Eqs.\ref{eq:Gauss_nd} and \ref{eq:simp_const_rel}, we arrive at the Poisson's equation as shown below. 
\begin{equation}
\label{eq:poisson}
n \cdot \left(\varepsilon_0 \varepsilon_r \mathrm{E}\right) = \rho_s
\end{equation}
Furthermore, the relationship between the electric field and electric potential is given by:
\begin{equation}
\label{eq:relation between E and V}
\mathrm{E} = -\nabla \mathrm{V}
\end{equation}
where $\nabla$ represents the gradient operator. Additionally, the gradiant and the curl of $\mathrm{E}$ are calculated by applying automatic differentiation with respect to the coordinates $\mathrm{X}$ to obtain the partial derivatives. Automatic differentiation is a family of techniques that evaluates derivatives and functions to compute gradients. Thus, so far, Eqs.\ref{eq:faraday}, \ref{eq:gauss}, \ref{eq:poisson}, and \ref{eq:relation between E and V} form the governing equations for the PINN framework.

Besides, the natural BCs are then given by: 
\begin{align} 
\mathrm{V}(\mathrm{X}) &= 0, X \in \Omega_{F,PCB} \label{eq_dirchilet_f_p}\\
\mathrm{V}(\mathrm{X}) &= 3.3, X \in \Omega_{sensor} \label{eq_dirchilet_s}
\end{align}
The above equations give the Dirichlet BCs defined over the coordinates $\mathrm{X}$ of finger $(\mathrm{F})$, PCB, and sensor. $\mathrm{V}(\mathrm{X})$ is the electric potential solution over the domains of interest $\Omega_{F}$, $\Omega_{PCB}$, and $\Omega_{sensor}$. The electric potential on the sensor is set to 3.3V based on the electronic circuit power usage. This can vary depending on the circuit used and typically ranges from 3.3V to 12V. $\mathrm{V}$ is set to zero on the finger and the PCB since they are grounded.
\paragraph{Neural Network Setup}
The central idea of PINN is to minimize a loss function in strong differential form given the governing equations and BCs. The loss minimization involves computing the loss function over a set of collocation points sampled from the conductors and the free-space domain. The inputs to the NN are four-dimensional, which includes spatial coordinates $\mathrm{X}$ and a finger distance $d$. The outputs of the network are the electric field distribution $\mathrm{E}=(\mathrm{E}_x,\mathrm{E}_y,\mathrm{E}_z)$, the electric potential $\mathrm{V}$, and the charge density $\rho$. The input collocation spatial coordinates are acquired primarily from two sources. A low-resolution simulation dataset $\Omega_S$ was acquired at three finger distances ($d_j$), namely, two extreme finger distances, $d_0=d_{min}$ and $d_1=d_{max}$, and one randomly chosen finger distance, $d_2=d_{m}$. Furthermore, a non-uniform 3D grid is designed by placing the finger-sensor setup at the center, over which the grid points are sampled. The distribution of this second source of grid points, $\Omega_G$, is refined in that they are more dense at the domain center and less dense at the outer edges. This allows focused utilization of the computational resources efficiently in areas with high variation. The sampled collocation points from the grid are merged with finger distances $d_h$ randomly sampled from a distance set $DS=[d_{min}...,d_{max}]$. At every iteration, collocation points from the two domains $\Omega_S$ and $\Omega_G$ are randomly sampled for training. 
\paragraph{Loss Functions}
The loss function for training the network, as shown in Eq. \ref{total_loss}, is divided into three components, namely $\pazocal{L}_{bc}$, $\pazocal{L}_{phy}$, and $\pazocal{L}_{data}$. $\pazocal{L}_{bc}$ refers to the essential and natural BC losses; $\pazocal{L}_{phy}$ indicates the physics losses computed over the governing PDEs; and $\pazocal{L}_{data}$ forms the final part of the combinatorial loss function computed from the low-resolution target data acquired from the simulation.  
\begin{align}
    \label{total_loss}
    &\pazocal{L}_{total} = \lambda_1 \pazocal{L}_{bc} + \lambda_2 \pazocal{L}_{phy} + \lambda_3 \pazocal{L}_{data}
\end{align}
The total loss $\pazocal{L}_{total}$ is a multi-part loss function with $\lambda_{i}>0, i \in (1,2,3)$ being the relative weighting coefficients. The individual loss functions are given by: 
\begin{equation}
\pazocal{L}_{bc} = \frac{1}{{{N_{bc}}}}\sum\limits_{i = 1}^{{N_{bc}}} {\left\| \mathrm{\hat{V}}_i - {\mathrm{V}_{{BC}_i}} \right\|^2} \label{l_bc} \\
\end{equation}
\begin{flalign}
\pazocal{L}_{phy} = \frac{1}{N_c}\sum\limits_{i = 1}^{N_c} \left( \left\| \mathrm{\hat{E}}_i + \nabla \mathrm{\hat{V}}_i \right\|^2  + \left\| - \nabla  \times \left( \varepsilon_0 \varepsilon_r \nabla \mathrm{\hat{V}}_i  \right) \right\|^2 +  \nonumber \right. \\
\left. \left\| \nabla  \cdot \left( \varepsilon_0 \varepsilon_r \mathrm{\hat{E}}_i \right) \right\|^2 + \left\| \hat{\rho}_i  - \left( \nabla  \cdot \left( \varepsilon_0 \varepsilon_r \mathrm{\hat{E}}_i \right) \right) \right\|^2   \right) \label{l_phy} \\
\pazocal{L}_{data} = \frac{1}{{{N_d}}}\sum\limits_{i = 1}^{{{N}_d}} \left( { {{\left\| {{\mathrm{\hat{E}}_i} - \mathrm{E}_i} \right\|}^2} + {\left\| {{\mathrm{\hat {V}}_i} - \mathrm{V}_i} \right\|}^2} + {\left\| {{{\hat \rho}_i} - {\rho_i}} \right\|}^2 \label{l_data}
\right)
\end{flalign}

where $\mathrm{\hat{E}}$, $\mathrm{\hat{V}}$, and $\hat \rho$ are the NN predicted electric field, electric potential, and charge density. 
$N_{bc}$, $N_c$, and $N_d$ refer to the number of collocation points sampled on the boundary, 3D domain, and data points, respectively. ${\left\|\cdot \right\|}^2$ denotes the mean square error (MSE) on the set of points.

$\pazocal{L}_{bc}$ corresponds to the BC losses computed over finger, sensor, and PCB between the NN outputs and the target condictions, where $\mathrm{V}_{BC_i}$ refers to the electric potential values for each element as shown in Eqs. \ref{eq_dirchilet_f_p} and \ref{eq_dirchilet_s}. A reference finger mesh pointcloud is generated at a specific distance, which is translated during the training iteration to extract the finger boundary points given the sampled query distance of the finger to the sensor. Sensor and PCB have fixed dimensions; thus, BCs are randomly sampled from their corresponding meshes during training to complete the total BC collocation input set. $\pazocal{L}_{phy}$ represents the residual physics losses computed from the governing PDEs. The relationship between the electric field and electric potential from Eq. \ref{eq:relation between E and V} is applied over all the coordinates sampled from the entire domains $\Omega_G$ and $\Omega_S$. $\pazocal{L}_{phy}$ is further composed of the residuals computed from the Faraday law (Eq. \ref{eq:faraday}) and the Laplace equation (Eq. \ref{eq:gauss}), where the collocation points in the interelectrode space are utilized. Finally, Poisson's equation (Eq. \ref{eq:poisson}) is used on the sensor and PCB coordinates using the NN predictions of surface charge density $\hat{\rho}_s$ and electric fields $\mathrm{\hat{E}}$. 
$\pazocal{L}_{data}$ refers to the data loss between the predicted $\mathrm{\hat{E}}$, $\mathrm{\hat{V}}$, $\hat \rho$ and simulation target dataset. 
The trained network is then evaluated for its accuracy using the normalized root mean square error (NRMSE), defined as 
\begin{equation}
{\text{NRMSE}_q} = \frac{{\sqrt{\frac{1}{M} \sum\limits_{i = 1}^M {{{\left\| {{{\hat q}_i} - {q_i}} \right\|}^2}} }}}{{ q_{max} - q_{min} }}
\end{equation}
where $q$ is the quantity of interest that comes from the simulation, $\hat{q}$ is the corresponding predicted quantity by NN, and $M$ is the total number of query points. Additionally, the mean absolute error (MAE) is also computed as a performance metric.

Based on the above loss functions, a fully connected neural network (FNN) with five hidden layers, each with 512 neurons with tanh activations, is used to train the PINN. Adam optimization is used in conjunction with the adaptive weight loss algorithm \citep{heydari2019softadapt} to train the model for 800k iterations. An initial learning of 1e-3 is used with an exponential decay. 
\section{\MakeUppercase{Experiments and Results}}

In this section, we evaluate the performance of the trained model. PDE-based loss minimization during training is conducted with 26 different finger distances to the sensor, ranging from a finger touching the sensor at $d_{min}=0$mm to a relatively far position of $d_{max}=25$mm. The value of relative permittivity $\varepsilon_r$ for the entire domain is set as 1. To evaluate the performance across this range, we aggregate the distances into three distinct proximity zones, namely \textit{near}, \textit{hover}, and \textit{far}. When a finger approaches a sensor, it disrupts the existing electric field generated by the sensor. The electric field concentration in the vicinity of the finger increases as the distance between them decreases. Aggregating the distances into three different zones allows one to get a better understanding of the performance of NN from a sensor functionality point of view. 

Table \ref{table:mae_and_mse} shows the MAE and NRMSE errors of the NN predictions. Charge density predicted by the NN is only used in Poisson's equation to minimize the PDE losses, whereas electric field distribution $\mathrm{E}$ and electric potential $\mathrm{V}$ are the high-relevant outputs that reflect the electrostatic characteristics. From the table, we can see that the NN exhibits reasonable generalization in all three proximity zones. The NRMSE values show that the maximum deviations in the predictions occur with the z-component of the electric field. These errors mainly occur in the immediate vicinity of the finger, where there is an abrupt change or a sharp transition in the electric field. This is more evident in the peak 25\% errors shown in the table. The peak 25\% errors are computed on the top 25\% of the maximum values for a given variable. This observation is in conjunction with \citep{sharma_2023}, who address this generalization issue that PINNs exhibit around boundaries. Furthermore, the high field gradients in the system around the conductive object also make it difficult for the NN to capture it accurately.
\begin{table*}[ht]
    \centering
    \captionsetup{justification=centering}
    \caption{Evaluation of MAE and NRMSE for Predicted NN Outputs E and V in the Three Proximity Zones}
    \label{table:mae_and_mse}
    \begin{tabular}{| c | c | c c c c | c c c c | }
        \hline
        \multirow{2}{*}{Mode} & \multirow{2}{*}{$d$ [mm]} & \multicolumn{4}{c|}{MAE (E [Volts/mm], V [Volts])}  & \multicolumn{4}{c|}{NRMSE}  \\ 
        \cline{3-10}
          &  & $\mathrm{E}_x$ & $\mathrm{E}_y$ & $\mathrm{E}_z$ & $\mathrm{V}$ & $\mathrm{E}_x$ & $\mathrm{E}_y$ & $\mathrm{E}_z$ & $\mathrm{V}$ \\
        \hline
        $near$ & $d \leq 2$  & 0.003 & 0.003 & 0.017 & 0.020  & 0.011  & 0.012 &	0.062 & 0.016 \\
        $hover$ & $2 < d \leq 10$ & 0.002 &	0.002 & 0.011 &0.015 & 0.009 &0.009 &0.047  &	0.012 \\
        $far$  & $10 < d \leq 25$ & 0.001 &	0.001 & 0.007 & 0.008 & 0.003 &0.004 &0.054  &	0.011 \\
        \hline
        \multicolumn{9}{c}{Peak 25\% Errors}\\
        \hline
        $near$ & $d \leq 2$	& 0.191 &	0.250 &	0.175 & 0.038 & 0.819 & 0.872 & 0.880  & 0.062 \\
        $hover$ & $2 < d \leq 10$  & 0.045 &	0.090 &	0.073 &0.024 	& 0.377 & 0.448 & 0.554 & 0.038\\
        $far$ &  $10 < d \leq 25$ & 0.035 &	0.018 &	0.019 & 0.008  & 0.260 & 0.222 & 0.322 & 0.014\\
        \hline
    \end{tabular}
\end{table*}
\begin{figure}[ht]
    \centering
    \includegraphics[width=0.45\textwidth]{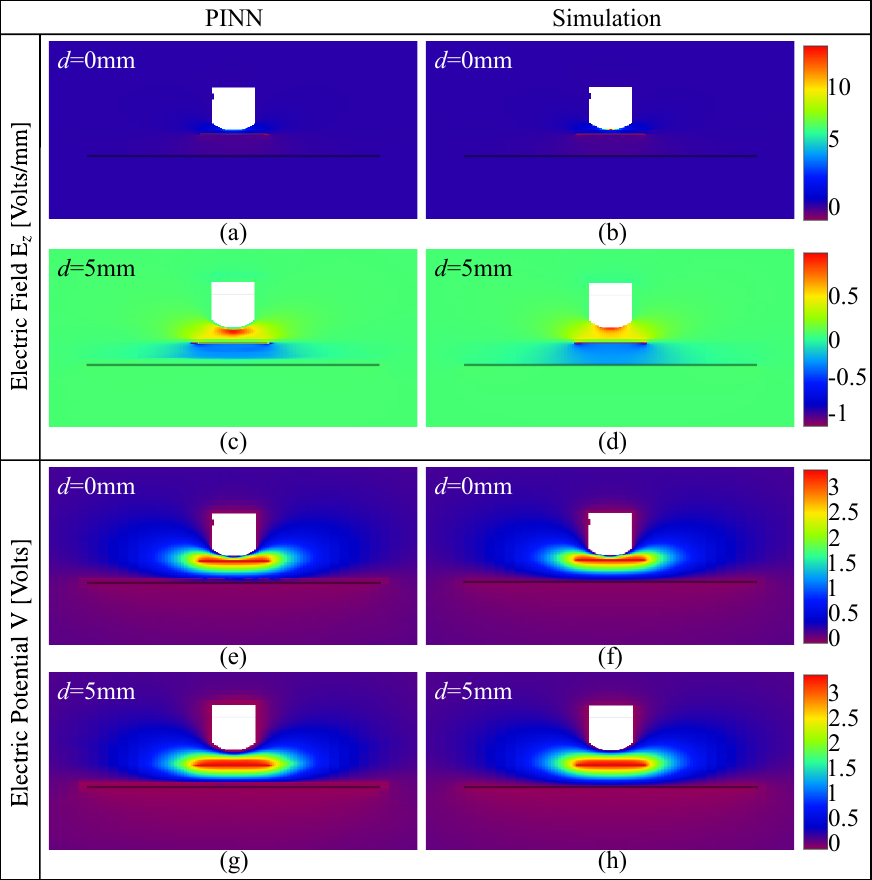}
    \captionsetup{justification=centering}
    \caption{Comparison of PINN Output and Simulation Target as Heatmaps at the Cross-section Plane $x=0$}
    \label{fig: 2d images plot}
\end{figure}

This phenomenon can also be seen in Fig. \ref{fig: 2d images plot}, which visualizes the heatmap of the predictions on the $yz$ plane slice at $x=0$ at two selected finger positions from proximity zones \textit{near} ($d=0$mm) (subplots:(a,b,e,f)) and \textit{hover} ($d=5$mm) (subplots (c,d,g,h)). The trained PINN is able to capture the physics quite good in a majority of the domain space with the highest error around the finger boundary edge. More complex and different PINN architectures, together with techniques such as exact enforcement of the Dirichlet BCs through signed distance functions, could improve the model accuracy \citep{berrone2023enforcing}. One of the future works arising out of this study will be focused around this subject. 

The PINN surrogate model demonstrates an overall 10x improvement in inference time compared to a traditional simulation tool. This does not account for the time expended on the experimental setup in the simulation but only the inference. Additionally, PINNs maintain their fast inference capabilities across different resolutions of input, further enhancing their utility in various applications. PINNs can leverage the computational power of GPUs, unlike existing traditional simulation tools, thereby enabling greater acceleration and efficiency in computational processing. Thus, PINNs exhibit a great potential to accelerate existing product development cycles and design optimization processes not just in the field of electrostatics but across a wide range of disciplines like fluid dynamics, structural mechanics, heat transfer, etc.

\section{\MakeUppercase{Conclusions}}
This paper presents a DNN-based PINN framework to solve the electrostatic interaction between a capacitive sensor and a finger by integrating Maxwell's equations. In particular, a scheme where the trained model predicts electrostatic characteristics such as electric potential and electric field distribution for a variety of finger positions with respect to the sensor is introduced. Spatial coordinates sampled from a non-uniform refined 3D grid along with randomly sampled finger distances are used as input to the NN. In addition, the model also predicts the charge density at a given point, which is directly utilized into the governing equation for loss minimization. Using a combination of data losses gathered from a sparse dataset at only three finger positions and a PDE-based physics-driven loss computed over a non-uniform refined 3D grid, a surrogate capacitive sensor model is trained successfully. The performance of the trained model is then evaluated against an unseen dataset generated from a multi-physics simulation. For this, we generated experimental data for 26 different finger positions to the sensor and organized them into three proximity zones. The results show great promise in its generalization capability at all distances, thereby establishing itself as a compelling surrogate sensor model. In the near future, our aim is to augment the current model through more advanced architectures and expand it for a capacitive sensor array.
\section*{\MakeUppercase{Acknowledgment}}

This work was supported in part by the Generalitat de Catalunya under the grant numbers 2021SGR01623 and 2022DI66.



\bibliography{refs}

\end{document}